
\input harvmac
\pretolerance=750
\def\etal{{\it et al.}}
\def\frc#1#2{{\textstyle{#1\over#2}}}

\def\prd#1#2#3{, {\sl Phys. Rev.} {\bf D#1} (19#2) #3}
\def\plb#1#2#3{, {\sl Phys. Lett.} {\bf #1B} (19#2) #3}

\def\SU#1{{\rm SU}(#1)}
\Title{HUTP-92/A057}{Trinification and the Strong $P$ Problem}
\centerline{Eric D. Carlson\footnote{$^*$}
{This research was supported by the National Science Foundation under Grant
\#\nobreak PHY-8714654, and the Texas National Research Laboratory
Commission under Grant \#RGY9206.}}
\medskip
\centerline{Meng Yuan Wang$^*$}
\bigskip\centerline{Lyman Laboratory of Physics}
\centerline{Harvard University}\centerline{Cambridge, MA 02138}

\vskip.3in
Models with spontaneously broken parity symmetry can solve the strong $CP$
problem in a natural way.  We construct such a model in the context of
$\SU3^3$ unification.  Parity has the conventional meaning in this model,
and the gauge group is unified.
\Date{6/92}

\newsec{Introduction}
One of the longstanding problems of particle physics is the so-called strong
$CP$ problem, namely, why is the coefficient $\theta$ so small in the
interaction
\eqn\brk{{\cal L} = {\theta g^2\over 64 \pi^2} \epsilon_{\mu\nu\alpha\beta}
C^{\mu\nu}_a C^{\alpha\beta}_a \; ,}
where $C^{\mu\nu}$ is the gluon field of chromodynamics, and
$\epsilon_{\mu\nu\alpha\beta}$ is the totally anti-symmetric tensor.  This
term respects charge conjugation invariance ($C$), but violates both parity
($P$) and their combination ($CP$).  When massive quarks are included, it
turns out that $\theta$ itself has no physical meaning, and can be redefined
by a chiral rotation of the quark fields.  However, the combination
\eqn\thetabar{\bar \theta = \theta + {\rm arg(\,det}(M_q))\; ,}
where $M_q$ is the quark mass matrix, is still physically meaningful, and
violates both P and CP.  Limits on the electric dipole moment of the neutron
imply \ref\thetalim{K.\ F.\ Smith \etal\plb{234}{90}{191}.}\ that $\bar\theta
\; \roughly{<} \; 2{\times}10^{-10}$.  The smallness of $\bar\theta$ is
conventionally called the strong $CP$ problem, but could just as well be
called
the strong $P$ problem.

One idea that has been pursued \ref\CPbreak{T.\ D.\ Lee\prd{8}{73}{1226}.}\ is
to have $CP$ as a spontaneously broken symmetry.  Parity can be used as well,
and Babu and Mohapatra built a model involving parity as a softly broken
symmetry.  Spontaneous parity violation is just as viable an option, as
pointed
out by Barr, Chang, and Senjnovi\'c \ref\Pbreak{S.\ M.\ Barr, D.\ Chang, and
G.\ Senjanovi\'c\prd{41}{90}{1286}}.  They pointed out that spontaneous parity
violation models cannot work without an extended gauge group.  Their model,
although it has many nice features, has the disadvantage that the symmetry
they call parity has nothing to do with conventional parity; it takes standard
model particles to heavy (as yet undiscovered) counterparts.

Our work differs from theirs in three ways.  First, parity affects standard
model particles the way you expect it to.  Second, we use the extended
gauge group to create a grand unified model.  Third, the whole context of
our model is built on a structure that has already been discussed in the
literature:  the trinification model of De R\'ujula, Georgi, and
Glashow \ref\Glash{A.\ De R\'ujula, H.\ Georgi, and S.~L.~Glashow, in: Fifth
Workshop on Grand Unification, eds.\ K.\ Kang, H.\ Fried, and P.\ Frampton
(World Scientific, Singapore, 1984)  p.\ 88.}.  In section 2 we will discuss
this unification scheme and how the idea of spontaneous parity breaking
applies to it.  In section 3 we will work out some of the phenomenological
consequences of this simple model.  In section 4 we will consider some simple
extensions of the idea.  In section 5 we will summarize our work.

\newsec{Trinification}
Trinification, one of many unification schemes, was originally promoted
\Glash\ for its relative simplicity of fermion and scalar content.  The
unification group is
\eqn\uni{G = \SU3_C \times \SU3_L \times \SU3_R \; ,}
where $\SU3_C$ is the standard color force, $\SU3_L$ contains the left-handed
$\SU2_L$ force of electroweak interactions, and the remaining ${\rm U}(1)$
part is distributed between $\SU3_L$ and $\SU3_R$.  In addition, there is a
cyclic $Z_3$ symmetry relating the three forces, so that they have the same
coupling constants.

Both the fermions and bosons appear in a 27 dimensional representation of $G$;
they transform under $G$ as the representation
\eqn\irrep{\Psi = \psi_L(3 , \bar 3, 1) + \psi_R(\bar 3, 1, 3) + \psi_\ell (1,
3, \bar 3)\; .}
The left-handed quarks and anti-right-handed quarks will be found in $\psi_L$
and $\psi_R$ respectively, and the leptons are found inside $\psi_\ell$.  The
scalars will acquire vacuum expectation values (VEV's) which are arranged as
\eqn\VEV{\left< \phi_\ell \right> = \pmatrix{u&0&0 \cr 0&u&u \cr 0 & w & v} \;
.}
The rows and columns of this matrix are understood to indicate the
transformation properties under $\SU3_L$ and $\SU3_R$ respectively, using the
same notation as \Glash.  The `constants' $u$, $v$, and $w$ here merely
represent orders of magnitude, rather than specific values.  The scales $v$
and $w$ break the symmetry group down to $\SU3_C \times \SU2_L \times \SU2_R
\times {\rm U}(1)_{B-L}$ and the standard model respectively, whereas $u$
accomplishes the electroweak breaking.  If there is only one scalar field, the
VEV can always be diagonalized, so that it is impossible to have the standard
model at an intermediate scale.  Hence it is necessary to assume at least two
27's, and, in the interest of economy, we will assume exactly two 27's.  These
fields suffice to perform all the necessary gauge symmetry breaking, both at
the unification scale and the electroweak scale.  To account for three
generations of fermions, however, we will assume three 27's of fermions.

We would like to consider promoting the simple $Z_3$ symmetry to the full
$S_3$ permutation group symmetry.  In other words, we would like to include
not only cyclic permutations of the three gauge fields, but pair switchings as
well.  The problem with this is that under these additional symmetries, the 27
does not transform into itself, it transforms into a $\overline{27}$.  This
suggests that we should also be including complex conjugation into our
symmetries.  However, we know that if $\Psi$ is a fermion field that
annihilates left-handed fermions, then $\Psi^*$ annihilates right-handed
fermions.  This implies that we must treat such transformations as parity
transformations.  Without further ado, let us write down the action of ${\bf
P}$, one of the three pair switching permutations, on all the fields:
\eqn\Pdef{\eqalign{
&\psi^A_\ell(\vec x, t) \to \psi^{A\dagger}_\ell(- \vec x, t)\; , \quad
\phi^i_\ell(\vec x, t) \to \phi^{i\dagger}_\ell(- \vec x, t)\; , \quad
C^\mu_a(\vec x, t) \to C_{a\mu} (-\vec x, t) \; ,\cr
&\psi^A_L(\vec x, t) \to \psi^{A\dagger}_R(- \vec x, t)\; , \quad
\phi^i_L(\vec x, t) \to \phi^{i\dagger}_R(- \vec x, t)\; , \quad
L^\mu_a(\vec x, t) \to R_{a\mu} (-\vec x, t) \; ,\cr
&\psi^A_R(\vec x, t) \to \psi^{A\dagger}_L(- \vec x, t)\; , \quad
\phi^i_R(\vec x, t) \to \phi^{i\dagger}_L(- \vec x, t)\; , \quad
R^\mu_{a}(\vec x, t) \to L_{a\mu} (-\vec x, t) \; ,\cr}}
where $C$, $R$, and $L$ are the gauge fields, the lowering of the index $^\mu$
to $_\mu$ indicates reversal of the spatial components, $A=1,2,3$ is a family
index and $i=1,2$ is a gauge boson index.  The daggers ($^\dagger$) represent
the
fact that not only are each component of these matrices complex conjugated,
but the $\SU3_L$ and $\SU3_R$ indices are exchanged as well.

${\bf P}$ is the action of one of the pair switchings on the fields; there are
also two other pair switchings which can be obtained by cyclic permutations of
${\bf P}$.  Note that the symmetry breaking at the scale $v$ does not break
${\bf P}$, though it does break the other two pair switchings.  This is why
${\bf P}$ corresponds to actual parity symmetry.

The symmetry ${\bf P}$ does not allow the parity breaking term \brk\ to
appear in the Lagrangian; hence it doesn't exist.  However, we must take more
care when considering the mass terms of the quarks; it is not immediately
obvious that there is no complex determinant to the quark mass matrix, and
hence a large $\bar\theta$.  This brings our attention to the Yukawa
couplings.

The Yukawa couplings responsible for the quark masses are given by
\eqn\Yuk{{\cal L}_{\rm Yuk} = - Z_3\{ f_{iAB} {\rm Tr}(\phi_\ell^i
\psi_L^A \psi_R^B) + h.c. \} \; ,}
where $Z_3$ simply implies that we must include cyclic permutations to assure
the $Z_3$ symmetry is respected.  Under the symmetry element ${\bf P}$, we can
relate the terms to their hermitian conjugates, so that $f_{iAB}^* = f_{iBA}$,
or, thinking of these as matrices, $f_i^\dagger = f_i$.  Assuming the scalar
VEV's are real, this will result in Hermitian quark mass matrices, or
$M_q^\dagger=M_q$.  Since the determinant of a Hermitian matrix is real, the
resulting $\bar\theta$ vanishes.

It remains only to confirm that the scalar VEV's are real.  To do so, we must
study the portion of the scalar potential responsible for the symmetry
breaking.  Since only the $\phi_\ell$ portions acquire VEV's, we will focus on
these.  The portion of the scalar potential that is relevant is given by
\def\pl{\phi_\ell}
\eqn\Ll{\eqalign{-{\cal L}_\ell = & m^2_{ij} {\rm Tr}(\pl^{i\dagger} \pl^j)
 + \{\gamma_{ijk} \epsilon_{\alpha\beta\gamma} \epsilon^{\delta\sigma\rho} \,
{\pl^i}_\delta^\alpha {\pl^j}_\rho^\beta {\pl^k}_\sigma^\gamma \; + \; h.c. \}
\cr
& + \lambda_{ijkl} {\rm Tr}(\pl^{i\dagger} \pl^j) {\rm Tr}(\pl^{k\dagger}
\pl^l)
+ \eta_{ijkl} {\rm Tr}(\pl^{i\dagger} \pl^j \pl^{k\dagger} \pl^l) \; . \cr}}

We now apply the parity symmetry together with the demand of Hermiticity to
show that, in fact, all of these constants are forced to be real.  Hermiticity
implies
\eqn\Hdemand{m_{ij}^2 = m_{ji}^{2*} \; , \quad \lambda_{ijkl} =
\lambda_{jilk}^* \; , \quad \hbox{and} \quad \eta_{ijkl} = \eta_{lkji}^* \; ,}
whereas parity implies
\eqn\Pdemand{m_{ij}^2 = m_{ji}^2 \; , \quad \gamma_{ijk} = \gamma_{ijk}^* \; ,
\quad \lambda_{ijkl} = \lambda_{jilk} \; , \quad \hbox{and} \quad \eta_{ijkl}
= \eta_{lijk} \; .}
It is obvious that these conditions together imply all the coefficients are
real with the possible exception of $\eta_{ijkl}$.  Using the equations
\Hdemand\ and \Pdemand, it is easy to show that all of the following are
equal:
\eqn\allequal{\eta_{ijkl}^* = \eta_{kjil} = \eta_{ilkj} = \eta_{jilk} =
\eta_{lkji} \; .}
The first two expressions force $\eta_{ijkl}$ to be real if $i=k$ or $j=l$.
Hence we need consider only when $i\ne k$ and $j \ne l$.  Because the indices
take on only the values 1 or 2, these conditions imply either $i=j$ and $k=l$
or $i=l$ and $j=k$.  Then the last two expressions assure that $\eta_{ijkl}$
is real, and thus the whole potential \Ll\ is real.  This makes it likely that
the VEV's will be real, resulting in Hermitian quark mass matrices and
vanishing $\bar \theta$, at least at tree level.  We will discuss $\bar
\theta$ at one loop level in the next section.

Thus we see that spontaneous parity violation in the context of trinification
provides a natural solution to the strong $CP$ problem.  It should be noted,
incidentally, that even though the potential involving only $\pl$ is real,
complex numbers occur both in the quark Yukawa couplings and in the scalar
potential in terms like ${\rm Tr}({\pl^1}^\dagger \pl^2 \phi_L^1
{\phi_R^1}^\dagger)$.  Thus the theory does not have $CP$ symmetry, only
parity.

\newsec{Phenomenology}
We now wish to do some more specific calculations.  To start with, let's
consider the three scales $u$, $v$, and $w$; we will assume the hierarchy $u
\ll w \ll v$.  At the unification scale $v$, the gauge couplings are all
equal, so that $\alpha_C = \alpha_L = \alpha_R$.  The electroweak interactions
are as strong as strong interactions and the weak angle is given by
$\sin^2\theta_W = \frc38$.  As we scale down first to $w$ and then to $u$, the
symmetry is broken down first to $\SU3_C \times \SU2_L \times \SU2_R \times
{\rm U}(1)_{B-L}$, then to the standard model, and finally to $\SU3_C \times
{\rm U}(1)_Q$.  Meanwhile, the coupling constants will evolve according to
the equation
\eqn\aevolve{{d\over dt}\left(4\pi \over \alpha_i\right) = \frc{22}{3} t_2(V)
- \frc83 t_2(F) - \frc23 t_2(S) \; ,}
where $t_2(V)$, $t_2(F)$, and $t_2(S)$ are the Casimir operators for the gauge
bosons, fermions, and scalars respectively, and $t$ is the logarithm of the
renormalization scale.

The exact results depend on the precise ways in which various masses appear at
each symmetry breaking.  The most naive assumption is that all particles that
acquire mass from a given symmetry breaking are immediately removed from
\aevolve.  Just above the electroweak scale $u$, there is no reason to assume
that any particles are light except for the three standard fermion families
and one Higgs doublet.  A right-handed neutrino and three more doublets must
be present above the scale $w$, and of course above the scale $v$ all
particles are massless.  Treating the onset of each scale as sudden, and using
the weak scale values $\alpha_C(M_Z) = 0.1134$, $\alpha_{em}(M_Z)
=\frc{1}{128}$, and $\sin^2\theta_W (M_Z) = 0.2325$, we find the approximate
unification scales
\eqn\unif{w = 4 {\times}10^{10}\,{\rm GeV} \qquad \hbox{and} \qquad v =
2{\times}10^{16}\, {\rm GeV} \; .}
Note that below the unification scale $v$, baryon number is conserved,
provided all of the colored scalar particles are very heavy.  Baryon number
violation proceeds through Yukawa couplings of the colored scalars, which
acquire masses at the scale $v$.  Because of the smallness of Yukawa couplings
and the largeness of $v$, proton decay limits are well within experimental
bounds.

It should be recognized that the scales in \unif\ are not precise.  In
particular, the heavy leptons acquire masses from the same Yukawa couplings
which are responsible for the light lepton masses.  The lightness of the
standard leptons may well be reflected in the lightness of these heavy
leptons, which will introduce thereshold effects at the high scales.  This
results in a substantial change in these scales to
\eqn\unifm{w = 1.2 {\times}10^{12}\,{\rm GeV} \qquad \hbox{and} \qquad v =
3{\times}10^{15}\, {\rm GeV} \; .}

The cancelling of $\bar \theta$ at tree level in models with spontaneous
symmetry breaking does not, in general, persist at higher loop levels.
In particular, loops of colored scalar particles will result in complex
effective quartic interactions involving two scalar fields transforming as
$(1,2,2,0)$ and two scalars transforming as $(1,2,1,{+}1)$ under the gauge
group $\SU3_C \times \SU2_L \times \SU2_R \times {\rm U}(1)_{B-L}$ below the
scale $v$.  When the VEV $w$ turns on, this will result in a complex mixing
among the fields transforming as $(1,2,2,0)$, some part of which is destined
to become the standard Higgs doublet.  This results in a phase mismatch
between the portion of the field coupling to the up-type quarks and the
portion
coupling to the down-type quarks, so that there will be a resulting phase in
the quark determinant and hence non-vanishing $\bar \theta$.  We consider this
a major problem with this theory in its simplest form.  It is, however,
impossible to estimate the magnitude of this contribution to $\bar  \theta$
because we know nothing about the quartic couplings in the original theory
which lead to this effect.

Most other contributions to $\bar \theta$ are either small or vanish.
Because $\bf P$ is still a good symmetry between the scales $v$ and $w$, $\bar
\theta$ vanishes to all orders above the scale $w$.  This causes most diagrams
involving colored scalars to be suppressed by two or more powers of $w/v$.
Contributions arising form the Yukawa couplings tend to be very
small and arise only at high loop level; indeed, we have not yet found any
nonvanishing contributions.  Basically, this occurs because it is possible to
use a vector unitary transformation to redefine the quark fields such that
only one phase appears among all the Yukawa couplings.  The situation is
exactly analogous to the standard model, where the Cabibbo-Kobayashi-Maskawa
(CKM) matrix has only one phase.

Finite neutrino masses are another interesting consequence of this theory.
At the scale $v$, the fields $\psi_{\ell 33}$ will acquire masses at one
loop level of the order $f^2 v/16\pi^2$, where $f$ is a generic quark-type
Yukawa coupling.  At the scale $w$ these heavy
neutrinos mix with $\psi_{\ell 32}$ with a fraction $w/v$, giving them
see-saw small masses of order $f^2 w^2/16\pi^2v$.  At the symmetry breaking
scale $u$, the lightest $\psi_{\ell 32}$ will acquire a Dirac mass of order
$hu$ connecting it with the standard neutrino $\psi_{\ell 23}$, where $h$
is a standard Yukawa coupling responsible for lepton masses.  This will,
by the see-saw mechanism, result in Majorana masses for the physical
neutrinos of order $16\pi^2 u^2vh^2/w^2f^2$.  Plugging in values from
\unifm, and guessing $h/f \approx m_\tau/m_t \approx 10^{-2}$, neutrino
masses of the order 1 keV are likely.  Such neutrino masses are possible
for the muon or tau neutrino, and could have very
interesting cosmological consequences.  Once again, it is impossible to
determine these masses with sufficient accuracy to make phenomenological
predictions.

\newsec{Extensions}
Several possible extensions of our work here seem worthy of note.  The
simplest and most obvious is to increase the number of scalar fields to match
the fermion fields.  Unfortunately, the constraints \Hdemand\ and \Pdemand\
are then no longer sufficient to assure the $\eta_{ijkl}$ is necessarily real.

Additional structure is required to avoid problems.  For example, if we impose
an additional $S_3$ symmetry among the families, where we simultaneously
permute the scalar and fermion family numbers, our solution is restored.

Seemingly more attractive is the idea of supersymmetrizing the theory, since
Higgs and fermions have the same $27$-dimensional representation. Also, our
desired VEV's, $v$ and $w$, seem to coincide with the F-flat direction of a
general superpotential. Upon close scrutiny, however, these nice features
evaporate quickly.  Although the F-terms are flat, the D-terms are positive
definite and therfore favor $v=w=0$.  To avoid this, it is necessary to
introduce additional $\overline{27}$'s to cancel these positive definite
terms.  Supersymmetric
$SU(3)^3$ models containing $n+3$ $27$'s and $n$ $\overline{27}$s have been
built for other reasons \ref\oxford{For example, n=6 in ``A Three-generation
Superstring Model'' by B.\ R.\ Greene, K.\ H.\ Kirklin, P.\ J.\ Miron and
G.\ G.\ Ross, Nuclear Physics B292(1987) 606.}, and do have a certain
number of pleasant features \oxford \ref\lighthiggs{M.~Y.\ Wang and E.~D.\
Carlson, ``Light Higgs without Fine-tuning'', in preparation}, but all the
phenomenological details have not been worked out.  Besides problems common
in other supersymmetric GUT's, the values of $v$
and $w$ remain controversial. Renormalization group calculation based on the
latest LEP data \ref\lep{U.Amaldi, W.de Boer and H.Furstenau, Physics Letters
B260(1991) 447} suggests that $v=w=10^{16.0\pm0.3}\,$GeV. It is difficult,
however, to obtain such large VEV's via the usual
soft-supersymmetry-breaking-versus-nonrenormalisable-term mechanism without
fine-tuning \ref\finetuning{P.Nath
and R.Arnowitt, Physical Review D39(1989) 2006}. Even if one is willing to
fine-tune, $v=w=10^{16}\,$GeV is not compatible with the fact that neutrino
masses are small \ref\neumass{R.Arnowitt and P.Nath, Physics Letters
B244(1990) 203.}. Furthermore, the phases of $v$ and $w$ are not determined,
and in general, can assume any values. Because these phases contribute
directly to $\bar{\theta}$, they will rule out all attempts to solve the
strong $CP$ ($P$) problem with spontaneous $CP$ ($P$) breaking.

All in all, the problems seem to proliferate just as fast as the solutions.

\newsec{Conclusion}
We have demonstrated the viability of $\SU3^3$ to explain the strong CP
problem in terms of spontaneous parity breaking.  Doubtless there are other
unification schemes which work as well.

Several problems remain, however.  The existence of wide disparities between
unification and electroweak scales (the hierarchy problem) is, of course,
still unexplained.  Spontaneous parity breaking shares with spontaneous CP
breaking the problem of domain walls coming from the breaking of discrete
symmetries \ref\domain{Ya.~B.~Zel'dovich, I.~Yu.Kobzarev, and L.~B.~Okun,
Sov.\ Phys.\ JETP 40 (1975) 1.}.  Inflation might well solve such problems
\ref\inflate{K.~Sato\plb{99}{81}{66}.}.  However, this problem existed in the
trinification model with only a $Z_3$ symmetry, so the extension to an $S_3$
symmetry does not necessarily make the problem any worse.

There is no explanation of why there should be three generations of fermions
and two of bosons.  Adding a third boson requires additional structure to
avoid the reintroduction of $\bar\theta$.  Supersymmetry has certain desirable
features when applied to this model, but the problems seem to proliferate
faster than their solutions.

It is difficult to make definite phenomenological predictions because of the
numerous parameters appearing in the theory.  However, it seems likely that
neutrino masses might lie in experimentally accessible regions.

We feel that our suggested solution to the strong CP problem deserves more
attention.  In particular, other unification schemes may also allow parity
symmetry with the possibility of spontaneous breaking and consequent vanishing
$\bar \theta$.  We hope this idea will be fully explored.
\vfill\eject
\listrefs
\bye